\documentclass[aps,prb,preprint,superscriptaddress,showpacs]{revtex4}

\begin{document}


\title{Ultrasonic Attenuation in the Vortex State of d-wave Superconductors}

\author{Tribikram Gupta }


\affiliation{Harish-Chandra Research Institute, Chhatnag Road,
 Jhusi, Allahabad 211019, India}

\author{D. M. Gaitonde}

\affiliation{High Pressure Physics Division, 
 Bhabha Atomic Research Centre, Mumbai 400085, India}


\begin{abstract}

We calculate the low temperature quasi-particle contribution to the ultrasonic attenuation rate 
 in the mixed state of d-wave superconductors. Our calculation is performed within the semi-classical approximation 
using quasi-particle energies that are Doppler shifted, with respect to their values in the Meissner phase, by the supercurrent associated with the vortices. We find that the 
attenuation at low temperatures and at fields $ H_{c1} \leq H \ll H_{c2}$ has
a temperature independent contribution which is proportional to $\surd H$ where $H$ is the applied magnetic field. We indicate how our 
result in combination with the zero-field result for ultrasonic attenuation can be used to calculate one of the parameters 
$v_F$, $H_{c2}$ or $\xi$ given the values for any two of them.

\end{abstract}


\maketitle

The discovery of high temperature superconductivity \cite {Bednorz} in the cuprates has led to an intense theoretical effort 
at understanding  the origins of the novel phenomena  seen in these materials. The normal state of the cuprates is highly 
anomalous \cite{PWA} and remains a puzzle that is still not understood. However, there is evidence \cite{Mohit} that well 
defined quasi-particles exist in the superconducting state. The superconducting state is now known \cite{Leggett, Kirtley} 
to be a d-wave state whose order parameter has $d_{x^2 - y^2}$ symmetry. The order parameter thus has nodes along the lines 
$k_x = \pm k_y$ in the two-dimensional Brillouin zone. 

While the origin  of the pairing interaction that leads to the occurence of d-wave superconductivity is not known at the present time, 
a great deal of progress in understanding the superconducting state can be made by focussing 
on the low energy nodal quasi-particles near the points   where the Fermi surface  intersects the  
$k_x = \pm k_y$ lines. These quasi-particles, which have an energy dispersion corresponding to Dirac fermions \cite{Lee} 
in the Meissner phase, are the dominant low-energy excitations  which determine the 
low temperature properties in the superconducting state. 
 
It is possible to derive \cite{MPLB} the d-wave gap from a model Hamiltonian but the connection of that  Hamiltonian 
with the underlying physics of strongly interacting fermions is unclear. Thus for the purposes of this paper, we assume on phenomenological grounds, the existence of BCS like quasi-particles with an energy gap that has d-wave symmetry. These 
quasi-particles might not be the real strongly interacting electrons, to whom their connection is unclear at the present time.

Ultrasonic attenuation has historically been a very useful tool in investigating the energies of quasi-particles 
in conventional superconductors. Early verification \cite{Morse} of the BCS prediction for the temperature dependence 
of the energy gap was done using  this technique. Even afterwards  \cite{Bishop,Batlogg} this has been a useful tool in the investigation 
of of superconductivity  in heavy fermion superconductors. On the theoretical front calculations of the ultrasonic attenuation 
in d-wave superconductors have been carried out both in the clean \cite{MPLB,Klemm,Vekh,Carb} and dirty \cite{Coleman, New} 
limit in the Meissner phase. The former limit which corresponds to $Ql \gg 1$ , Q being the ultrasound wave-vector and l being the electronic mean free path, is the one considered by us in this paper.

Assuming 100 MHz to be a typical frequency \cite{Bishop,Batlogg} (although ultrasound experiments can be carried out over a large 
frequency range from kHz to GHz) at which ultrasound experiments are done and taking the sound velocity to be\cite{Shobo} $4\times{10^5}$m/s we find that this translates into the requirement $l\gg 4\mu m$. This restricts the applicability of 
our work to the cleanest samples\cite{Clean}.

While ultrasonic attenuation in the vortex state has recieved attention in the past \cite{Pankert,Faraday1, Faraday2,Blatter} the 
emphasis of earlier workers has been to study  effects arising from the coupling of ultrasonic waves to  vortex motion. 
This coupling comes from two different effects: a) The pinning potentials from impurity ions tend to drag the vortices 
along with them , b) The ionic current due to the ionic displacement  will exert a Lorentz force on the vortices. In 
the clean limit($l \geq {4\mu-m}$) in which we work, the effects of pinning are expected to be weak. The Lorentz force 
coupling is in any case weak because it is suppressed by a factor of $v/c$ where $v$ is the ionic velocity and $c$ 
is the velocity of light. We therefore ignore these processes and focus entirely on the  phonon damping from the 
electronic quasi-particles.

In conventional s-wave superconductors, low energy electronic excitations  in the vortex state are the bound states localised 
in the  core region of the vortices \cite{Caroli}. However in d-wave superconductors, because of the presence of nodes in the gap
 function, the dominant low energy excitations are those in the far region away from the vortex cores. Their contribution is expected to be overwhelmingly larger \cite{Volovik}  than that coming from the  the cores. Further, experimental 
studies from STM measurements \cite{STM1,STM2} reveal that that there are just a few bound states in the vortex cores. Thus we focus 
exclusively on the excitations in the far region.

To describe the low energy excitations in the mixed phase we make the use of the semi-classical approximation, first discussed \cite{DeGennes} in the case of s-wave superconductors. This approximation has been employed for d-wave superconductors  in recent  
times \cite{Volovik,Hirschfeld} and its regime of applicability is the range $H_{c1} \leq H \ll H_{c2}$, where H is the applied 
magnetic field and $H_{c1}$ and $H_{c2}$ are the lower critical and upper critical magnetic fields repectively. A striking success 
of this method has been the prediction \cite{Volovik} of a term linear in the temperature T whose co-efficient scales as $\surd H$
(H being the applied magnetic field). This prediction has been verified \cite{spht} and
has given greater credibility to the semi-classical description. Recently this approximation has
been systematized and put on a firm footing by  Ramakrishnan and 
Rajagopal\cite{TVR} who have
derived it microscopically.

Using the semi-classical approximation to describe the electronic Green functions in the superconducting state we calculate the
imaginary part of the electron density-density correlation function which is proportional to the inverse phonon lifetime. As 
the sound velocity is only weakly dependent on temperature\cite{Shobo}, the temperature dependence of the attenuation comes almost entirely 
from the the inverse lifetime. We find that at low tempeartures, and for parameters appropiate to the cuprates,
the ultrasonic attenuation co-efficient in the mixed state has a temperature independent contribution which scales as $\sqrt H$. The 
co-efficient of this term and its dependence on the ultrasound wave-vector as well as the leading temperature corrections to it have 
been explicitly evaluated. These are our main results.

We now present the details of our calculation. The semi-classical approximation can be  understood \cite{DeGennes}as ignoring, 
in the first instance, the spatial variation of the supercurrents around a vortex outside its core region. This can be justified 
on the grounds that the spatial variation of the electronic wave-function has a characteristic length scale $k_F^{-1}$ which is 
much smaller than the smallest length scale associated with vortex currents $\xi$ (the core radius). In the cuprates the 
parameter $(k_F\xi)^{-1}\simeq 10$ and thus the semi-classical approximation is expected to provide a reasonable description. 
We then proceed to evaluate the inverse phonon lifetime to first order in ${\vec q}$
where a current has been introduced by taking the energy gap function to be $\Delta_{\vec q}({\vec{R_i}},{\vec{R_j}}) = 
\Delta_{i,j}e^{i{\vec q}\cdot({\vec{R_i}} + {\vec{R_j}})}$. Here ${\vec{R_i}}$ and ${\vec{R_j}}$ are neighbouring sites on a square 
lattice whose fermions are paired up in a  singlet state and $\Delta_{i,j} = \Delta(-\Delta)$ for ${\vec{R_i}} - {\vec{R_j}}$ 
being in the $\hat x(\hat y)$ direction. The spatial dependence of ${\vec q} = \hat{\phi}/2r$ is now restored and the the inverse 
phonon lifetime is averaged over a unit-cell of the vortex lattice to get our final result for the attenuation co-efficient.

Solving the Bogulibov-deGennes equation for the mean-field d-wave superconductor to linear order in ${\vec q}$ we obtain the results:
\begin{equation}
G_{\vec q}^{\sigma}({\vec{R_i}} - {\vec{R_j}},i\omega_n) = {1\over N}{\sum_{\vec k}}{e^{i{\vec k}\cdot({\vec{R_i}} - {\vec{R_j}})}}
[{u_{{\vec {k}} -{\vec {q}}}^2\over{i\omega_n - E_{{\vec {k}} - {\vec {q}}}}} + {v_{{\vec {q}} -{\vec {k}}}^2\over{i\omega_n + E_{{\vec {q}} - 
{\vec {k}}}}}]
\end{equation}
where $ {u_{\vec k}}^2 = (1 + \epsilon_{\vec k}/E_{\vec k}^{0})/2, {v_{\vec k}}^2 = (1 - \epsilon_{\vec k}/E_{\vec k}^{0})/2, \epsilon_{\vec k} = \xi_{\vec k} - \mu, \xi_{\vec k}$ being the 
band energy and $\mu$ the chemical potential, $E_{\vec k}^0 = \sqrt{{\epsilon_{\vec k}^2} + {\Delta_{\vec k}^2}}, 
\Delta_{\vec k} = \Delta( \cos{k_{x}a} - \cos{k_{y}a})$ being the d-wave gap and $E_{\vec k} = E_{\vec k}^0 + {\vec q}\cdot\nabla_k\xi_{\vec k}$ 
being the Doppler shifted quasi-particle energy. Here

\begin{equation}
G_{\vec q}^{\sigma}({\vec{R_i}} - {\vec{R_j}},i\omega_n) = -\int_0^{\beta}d\tau<c_{i,\sigma}(\tau)c_{j,\sigma}^{\dagger}(0)>e^{i\omega_n\tau}     
\end{equation}
is the "normal" Green function.
Similarly the anomalous Green functions are given by:

\begin{equation}
F_{\vec q}({\vec{R_i}},{\vec{R_j}},i\omega_n) = {e^{i{\vec q}\cdot({\vec{R_i}} + {\vec{R_j}})}\over N}\sum_{\vec k}
{u_{\vec k}v_{\vec k}} 
({e^{-i{\vec k}\cdot({\vec{R_i}} - {\vec{R_j}})}\over{i\omega_n - E_{\vec k}}} -  
{e^{i{\vec k}\cdot({\vec{R_i}} - {\vec{R_j}})}\over{i\omega_n + E_{\vec k}}})      
\end{equation}

and

\begin{equation}
F_{\vec q}^{+}({\vec{R_i}},{\vec{R_j}},i\omega_n) = {e^{-i\vec q\cdot(\vec{R_i} + \vec{R_j})}\over N}\sum_{\vec k}
{u_{\vec k}v_{\vec k}} 
({e^{-i\vec k\cdot(\vec{R_i} - \vec{R_j})}\over{i\omega_n - E_{\vec k}}} -  
{e^{i\vec k\cdot(\vec{R_i} - \vec{R_j})}\over{i\omega_n + E_{\vec k}}})      
\end{equation}
 
Here $u_{\vec k}v_{\vec k} = \Delta_{\vec k}/2E^{0}_{\vec k}$ and the Green functions $F_{\vec q}$ and $F_{\vec q}^{+}$ are defined as 
\begin{equation}
F_{\vec q}(\vec{R_i},\vec{R_j},i\omega_n) = -\int_0^{\beta}d\tau<c_{i,\downarrow}(\tau)c_{j,\uparrow}(0)>e^{i\omega_n\tau}     
\end{equation}

\begin{equation}
F_{\vec q}^{+}(\vec{R_i},\vec{R_j},i\omega_n) = -\int_0^{\beta}d\tau<c_{i,\uparrow}^{\dagger}(\tau)c_{j,\downarrow}^{\dagger}(0)>e^{i\omega_n\tau}     
\end{equation}

Using the Green functions in equations (1),(3) and (4) we compute the imaginary part of the density-density correlation function which is given by

\begin{equation}
\chi^"_{\vec q}(\vec Q,\omega) = X_1 + X_2 + X_3 + X_4
\end{equation}      

where 

\begin{equation}
X1= {-2\pi\over N}\sum_{\vec k}[n(E_{\vec k})-n(E_{{\vec k}+{\vec Q}})]
({u^2_{{\vec k}+{\vec Q}}}u^2_{\vec k} - {\Delta_{\vec k}\Delta_{{\vec k}+{\vec Q}}\over 4 E_{\vec k}^0
E_{{\vec k}+{\vec Q}}^0})\delta(\omega+E_{\vec k}-E_{{\vec k}+{\vec Q}})
\end{equation}

\begin {equation}
X2={-2\pi\over N}\sum_{\vec k}[n(E_{-\vec k})-n(E_{{-\vec k}{-\vec Q}})]
(v_{{\vec k}+{\vec Q}}^2v_{\vec k}^2-{\Delta_{\vec k}\Delta_{{\vec k}+{\vec Q}}\over 4 E_{\vec k}^0
E_{{\vec k}+{\vec Q}}^0})\delta(\omega-E_{-\vec k}+E_{{-\vec k}-{\vec Q}})
\end{equation}

\begin{equation}
X3={-2\pi\over N}\sum_{\vec k}[1-n(E_{-\vec k})-n(E_{{\vec k}+{\vec Q}})]
(u_{{\vec k}+{\vec Q}}^2v_{\vec k}^2+{\Delta_{\vec k}\Delta_{{\vec k}+{\vec Q}}\over 4 E_{\vec k}^0
E_{{\vec k}+{\vec Q}}^0})\delta(\omega-E_{-\vec k}-E_{{\vec k}+{\vec Q}}),
\end{equation}

and 

\begin{equation}
X4={2\pi\over N}\sum_{\vec k}[1-n(E_{\vec k})-n(E_{-{\vec k}-{\vec Q}})]
(v_{{\vec k}+{\vec Q}}^2u_{\vec k}^2+{\Delta_{\vec k}\Delta_{{\vec k}+{\vec Q}}\over 4 E_{\vec k}^0
E_{{\vec k}+{\vec Q}}^0})\delta(\omega+E_{\vec k}+E_{-{\vec k}-{\vec Q}}).
\end{equation}

In s-wave superconductors the contribution of $X_3$ and $X_4$ are zero as the ultrasound frequency $\omega \ll 2\Delta $ and so the 
$\delta$-fn condition in Eqs.(10) and (11) can never be satisfied. In the d-wave case the presence of nodes in $\Delta_{\vec k}$
means that $X_3$ and $X_4$ will be finite. However, as the phase space for them is limited, their contribution is expected to be small 
and so we focus exclusively on $X_1$ and $X_2$ which make the dominant contribution to $\chi^"_{\vec q}(\vec Q,\omega)$.

Now expanding to the leading order in $\omega$ we find 

\begin{eqnarray}
\chi^"_{\vec q}(\vec Q,\omega) = {2\pi\omega\over N}\sum_{\vec k}[n'(E_{\vec k})
\delta(E_{\vec k} - E_{\vec {k} + \vec {Q}})
(u^2_{\vec {k} + \vec {Q}}u^2_{\vec k} - {\Delta_{\vec k}\Delta_{{\vec k}+{\vec Q}}
\over{4 E_{\vec k}^0E_{{\vec k}+{\vec Q}}^0}})  
+ n'(E_{-\vec k}) \nonumber \\
\delta(E_{-\vec k} - E_{-{\vec k}-{\vec Q}}) 
(v^2_{{\vec k}+{\vec Q}}v^2_{\vec k} - {\Delta_{\vec k}\Delta_{{\vec k}+{\vec Q}}
\over {4 E_{\vec k}^0 E_{{\vec k}+{\vec Q}}^0}})]
\end{eqnarray}

We next expand   $\chi^"$ to leading order in the ultrasound wave-vector $\vec Q$ and on making the substitution $\vec {k}\rightarrow \vec {-k}$ in the second term 
in Eq. (12) we arrive at the result

\begin{equation}
\chi^"_{\vec q}(\vec Q,\omega)= {2\pi\omega\over N}\Sigma_{\vec k}n'(E_{\vec k}^0 + \vec {q}\cdot\vec{\nabla_k}\xi_{\vec k})
{\epsilon_{\vec k}^2\over E_{\vec k}^{0^2}}\delta(\vec {Q}\cdot\vec {\nabla_k}E^0_{\vec k} + q_{\alpha}Q_{\beta}{{\partial^{2}\xi_{\vec k}}\over{{\partial{k_{\alpha}}}{\partial{k_{\beta}}}}})
\end{equation}
where a sum over repeated indices is implied in the arguement of the $\delta$ function.

To do the ${\vec k}$ summation in Eq. (13) we only consider \cite{Lee} the fermions near the nodes. This approximation is reasonable at 
low temperatures $(k_{B}T \ll \Delta)$. For the node near $\vec {P_a} = (\pi/{2a} , \pi/{2a})$ upon linearising the band 
energies and gap function, we have: $\xi_{\vec k} \simeq v_{F}k_1$ and $\Delta_{\vec k} \simeq v_{\Delta}k_2$ where $k_1$ and 
$k_2$ are co-ordinates normal and tangential respectively to the Fermi surface at the node. In terms of the co-ordinates 
$k_1$ and $k_2$ we find that the contribution from this node to $\chi^{"(a)}_{\vec q}$ can be written as

\begin{eqnarray}
\chi^{"(a)}_{\vec q}(\vec Q, \omega) \simeq {{a^2\omega}\over{2\pi}}
\int{dk_1}\int{dk_2}
n'[\sqrt {v_{F}^2k_1^2 + v_{\Delta}^2k_2^2}  + \rho_0]
{v_{F}^2k_1^2\over{v_{F}^2k_1^2 + v_{\Delta}^2k_2^2}} \nonumber \\ 
\delta[{{{\alpha} v_{F} k_1 + {\beta} v_{\Delta} k_2}\over {\sqrt{v_{F}^2k_1^2 + v_{\Delta}^2k_2^2}}} 
- \alpha_{1}v_{F}k_1 - \alpha_{2}v_{F}k_2]   
\end{eqnarray}

Here $\alpha=v_FQ\cos(\pi/4-\theta)$, $\beta=v_{\Delta}Q\sin(\pi/4-\theta)$,
$\alpha_1 = (Qqa^2/2)\cos(\psi - \theta)$,
$\alpha_2 = -(Qqa^2/2)\cos(\psi + \theta)$, $\rho_0=qv_{F}\cos(\psi - \pi/4)$ and $\theta$ and $\psi$ are the angles made by 
$\vec Q$ and $\vec q$ respectively with respect to the $k_1$ axis.
We now introduce the polar co-ordinates $v_{F}k_1$ = $\rho{\cos{\phi}}$ and $v_{\Delta}k_2$ = $\rho{\sin{\phi}}$. 
Then upon performing the $\phi$ integral we find $\chi^{"(a)}_{\vec q}$ reduces to 

\begin {equation}
\chi^{"(a)}_{\vec q}({\vec Q}, \omega) = {a^2\omega\over {{\pi}v_Fv_{\Delta}}} {\int}_{0}^{\rho_c}d\rho\rho
n'[\rho + \rho_0]f_a[\rho]
\end {equation}

where $f_a[\rho] = {B^2\over (A^2 + B^2)^{3/2}}$ and  $A = \alpha - \alpha_{1}\rho$ , 
$B = \beta - (v_F/v_{\Delta})\alpha_{2}\rho$ and $\rho_c = \sqrt {{\pi}v_{F}v_{\Delta}}/a$ 
is a cutoff introduced to preserve the volume of the Brillouin zone while doing 
the ${\vec k}$-integration.

Introducing the variable $\rho' = \rho + \rho_0$ we have $\chi^{"a}=\chi^{"(a1)}+
\chi^{"(a2)}$ where
\begin {equation}
\chi^{"(a1)}_{\vec q}(\vec Q, \omega) = {a^2\omega\over {{\pi}v_Fv_{\Delta}}} {\int}_{\rho_0}^{\rho_c}
d\rho'{\rho'}n'[\rho']f_a[\rho' - \rho_0]
\end {equation}
and
\begin {equation}
\chi^{"(a2)}_{\vec q}(\vec Q, \omega) = -{a^2\omega\over {{\pi}v_Fv_{\Delta}}}\rho_0 {\int}_{\rho_0}^{\rho_c}
d\rho' n'[\rho']f_a[\rho' - \rho_0]
\end {equation}
In writing Eqs. (16) and (17) we have made use of $\rho_c\gg\rho_0$ which follows from using
$q_{max}=1/2\xi$ and $\xi\approx 30\AA$, $v_F/a\approx 192meV$ and $v_{\Delta}/a\approx 28meV$
for parameters appropriate to the cuprates\cite{Lee}.

In an analogous fashion we evaluate the corresponding contributions to $\chi^"_{\vec q}$ from 
the nodes near $\vec {P_b} = (-\pi/{2a} , -\pi/{2a})$, $\vec {P_c} = (-\pi/{2a} , \pi/{2a})$ 
and $\vec {P_d} = (\pi/{2a} , -\pi/{2a})$. We first focus on the type of terms in Eq. (16). In 
this case we find, on expanding to linear order in $q$, that the contribution linear in $q$ 
exactly vanishes due to cancellations from the terms coming from different nodes. We thus find 
$\chi^{"(1)}_{\vec q} = \chi^{"(a1)}_{\vec q} + \chi^{"(b1)}_{\vec q} + \chi^{"(c1)}_{\vec q} 
+ \chi^{"(d1)}_{\vec q}$ to be given by

\begin{equation}
\chi^{"(1)}_{\vec q} = -{{2ln2{\omega}{a^2}k_{B}T}\over{{\pi} v_{F} v_{\Delta}}}
[{\beta^2\over {(\alpha^2 + \beta^2)^{3/2}}} + {\eta^2\over {(\lambda^2 + \eta^2)^{3/2}}}]
\end{equation}

to linear order in $\vec q$. Once again we have used $k_{B}T \ll qv_F \ll \rho_c$.
Here $\alpha$ and $\beta$ are previously defined and we have introduced the parameters $\lambda=v_FQ\sin(\pi/4-\theta)$, $\eta=v_{\Delta}Q\cos(\pi/4-\theta)$. 
The result obtained for this term is identical to the zero field result for the attenuation 
previously calculated in Ref. (12). We next turn our attention to the terms of the type written 
in Eq.(17). Here on adding the contributions from the nodes near $\vec P_a$ and $\vec P_b$, we 
find to leading order in $q$,
\begin{equation}
\chi^{"(a2)}_{\vec q} + \chi^{"(b2)}_{\vec q} = -{{\omega}{a^2}q\over {{\pi}v_{\Delta}}}
\cos(\psi - \pi/4){\beta^2\over {(\alpha^2 + \beta^2)^{3/2}}}\tanh({qv_F\cos(\psi-\pi/4)\over 2k_BT})
\end{equation}
Similarly the nodes near $\vec P_c$ and $\vec P_d$ yield the contribution
\begin{equation}
\chi^{"(c2)}_{\vec q} + \chi^{"(d2)}_{\vec q} = -({{\omega}{a^2}q\over \pi v_{\Delta}})
\cos(\psi + \pi/4){\eta^2\over {(\lambda^2 + \eta^2)^{3/2}}}\tanh({qv_F\cos(\psi+\pi/4)\over 2k_BT})
\end{equation}

It is straightforward to see that for $T\rightarrow 0$, Eqs. (19) and (20) reduce to
\begin{equation}
\chi^{"(a2)}_{\vec q} + \chi^{"(b2)}_{\vec q} = -({{\omega}{a^2}q\over {{\pi}v_{\Delta}}})
\cos(\psi - \pi/4){\beta^2\over {(\alpha^2 + \beta^2)^{3/2}}}[\theta(\cos(\psi-\pi/4)) 
- \theta(-\cos(\psi-\pi/4))]
\end{equation}

and 

\begin{equation}
\chi^{"(c2)}_{\vec q} + \chi^{"(d2)}_{\vec q} = -({{\omega}
{a^2}q\over {{\pi}v_{\Delta}}})
\cos(\psi + \pi/4){\eta^2\over {(\lambda^2 + \eta^2)^{3/2}}}
[\theta(\cos(\psi+\pi/4)) - \theta(-\cos(\psi+\pi/4))]
\end{equation}

Eqs.(21) and (22) are also obtained directly by taking the T = 0 limit for n' in Eq.(15) and its analogues for the other 3 nodes.

To proceed further we now restore 
$\vec q = {\hat{\phi}}/{2r}$ corresponding  to the supercurrents around a 
vortex where the vector potential has been ignored as the Ginzburg-Landau parameter $\kappa \gg 1$. 
Assuming a vortex lattice with circular unit-cells 
we find that the radius of the cells $R_c \sim  \xi \sqrt{H_{c2}/H}$. 
We then average
$ \chi^"_{\vec q}$ over one unit cell to obtain  
\begin{equation}
\chi^{"}(H,T) = {{\int_{\xi}^{R_c}rdr\int_0^{2\pi}d\psi\chi^"_{\vec q}}
\over{\int_{\xi}^{R_c}rdr\int_0^{2\pi}d\psi}}
\end{equation}
to obtain our final result for $\chi^"$. 

In order to perform the averaging in Eq. (23) we make the approximation 
$tanh(x) \simeq x$ for
$|x|<1$, $tanh(x) \simeq 1$ for $x>1$ and $tanh(x) \simeq -1$ for $x<-1$.
This approximation interpolates between the asymptotically exact behaviour at
small and large $\mid x\mid$. Further it reproduces the exact result that can be directly
calculated for $T=0$.  
In that case the integral 
in Eq. (23)  is elementary and can be done by substituting the expressions 
in Eqs. (21) and (22). We then obtain to leading order in the small parameter 
$(\xi k_BT/v_F)^2(H_{c2}/H)$:

\begin{equation}
\chi^{"(2)}(H,T) = -{2{\omega}a^2\over {{{\pi}^2}{v_{\Delta}}{\xi}}}
({H\over H_{c2}})^{1/2}[1 - {8\xi^2\over 9}({k_{B}T\over v_F})^2{H_{c2}\over H}] 
[{\beta^2\over{(\alpha^2 + \beta^2)^{3/2}}} + 
{\eta^2\over {(\lambda^2 + \eta^2)^{3/2}}}]
\end{equation}

The assumption about the small parameter implies that the regime of 
validity of our result is for
$k_{B}T \ll {v_{F}\over {\xi}}({H\over H_{c2}})^{1/2}$ together with the condition 
$H_{c1} \leq H \ll H_{c2}$. 
For parameters relevant to the cuprates $H_{c2}/H_{c1} \simeq 100$ at low temperatures. 
This restricts the temperature 
window to $T \ll 1K$ at $H = H_{c1}$. 
Our result for a temperature independent contribution to the attenuation that scales as $\sqrt{H}$ is 
entirely understandable as it has its origin in a finite density of states at the Fermi-energy \cite{Volovik} whose size is proportional to ${\sqrt H}$ and
whose signature is seen in the specific heat measurements\cite{spht}.

The actual ultrasonic attenuation 
coefficient is 
related to $\chi^"$ by 
\begin{equation}
\alpha_S(T,H) = M(\vec Q)\chi^"(\vec Q, T, H)
\end{equation}
where $M(\vec Q)$ is a constant which depends on the sound velocity and 
the electron-phonon matrix 
element and the ultrasound frequency. Then on combining Eqs. (18) and (24) 
and considering the fact 
that Eq. (18) contains the result for the Meissner phase ($H=0$) we obtain the result:
\begin{equation}
{{\alpha_{S}(T,H) - \alpha_{S}(T,H=0)}\over{\alpha_{S}(T,H=0)}} = {v_{F}\over{{\pi}{\xi}}}({H\over H_{c2}})^{1/2}
{1\over{ln2k_{B}T}}[1 - {{8{\xi}^2}\over9}({{k_{B}T}\over{v_F}})^2{H_{c2}\over H}]
\end{equation} 
whose window of validity has been described above. This result is remarkable because it is independent
of the ultrasound wave-vector ${\vec Q}$ as well as the gap increase parameter $v_{\Delta}$. Thus a measurement 
of the ultrasonic attenuation at low temperatures and at fields above $H_{c1}$ would enable a determination of 
any one of the parameters $v_F$,$\xi$ and $H_{c2}$ given a knowledge of the other two.

We now discuss some shortcomings of our work. We have assumed perfectly well-defined quasi-particles thus 
ignoring the incoherent spectral weight seen in photoemission experiments \cite{Mohit}. The semi-classical 
approximation used by us may not accurately describe all the physical effects due to the scattering of 
quasi-particles from the supercurrents  in the vortex state. Our results are restricted to the clean limit
and are applicable only to very clean samples.

Finally we conclude by recapitulating the main points of this paper. We have employed the semi-classical
approximation to evaluate the phonon damping due to the electronic quasi-particles in a d-wave superconductor. 
We find that for parameters appropiate to the cuprates, in a temperature window $k_{B}T \ll 1K$, there is a 
temperature independent contribution to the ultrasonic attenuation whose magnitude scales as $\sqrt{H}$. 

\begin{acknowledgments}
One of us (DMG) thanks T.V.Ramakrishnan for useful comments.
\end{acknowledgments}


\begin{thebibliography}{99}
\bibitem {Bednorz} J. G. Bednorz and K. A. Muller, {\it Z. Phys.}{\bf B64}, 189 (1986).
\bibitem {PWA} P. W. Anderson, {\it The Theory of Superconductivity in High-Tc Cuprates},
(Princeton University Press, Princeton, 1997).
\bibitem {Mohit} M. Randeria et. al, {\it Phys. Rev. Lett.} {\bf 74}, 4951 (1995). 
\bibitem{Leggett} J.F.Annett, N.Goldenfeld and A.J.Leggett, in {\it Physical Properties of High Temperature Superconductors},
{\bf Vol-5}, ed. D.M.Ginsberg (World Scientific, Singapore,1996), p. 375.
\bibitem{Kirtley} C. C. Tsuei and J. R. Kirtley, {\it Rev. Mod. Phys.} {\bf 72}, 969 (2000).
\bibitem{Lee} P.A.Lee and X.G.Wen, {\it Phys. Rev. Lett.} {\bf 78}, 4111 (1997).
\bibitem{MPLB} T.Gupta and D.M.Gaitonde, {\it Mod.Phys.Lett.} {\bf B15}, 269 (2001).
\bibitem {Morse} R. W. Morse in {\it Progress in Cryogenics},
{\bf Vol-I}, ed. K. Mendelssohn (Heywood, London, 1959) p.219.
\bibitem{Bishop}D. J. Bishop et. al., {\it Phys. Rev. Lett.}, {\bf 53}, 1009 (1984).
\bibitem{Batlogg} B. Batlogg et. al., {\it Phys. Rev. Lett.}, {\bf 55}, 1319 (1985).
\bibitem{Klemm} S.N. Coppersmith and R.A. Klemm, {\it Phys. Rev. Lett.}{\bf 56}, 1870(1986).
\bibitem {Vekh} I.Vekhter, E.J. Nicol and J.P. Carbotte, {\it Phys. Rev.} {\bf B59}, 7123 (1999).
\bibitem {Carb}W.C.Wu and J.P.Carbotte, {\it Phys. Rev.} {\bf B60}, 14943 (1999).
\bibitem{Coleman} J.Moreno and P.Coleman, {\it Phys. Rev.} {\bf B53}, R2995 (1996).
\bibitem{New}M.B.Walker, M.F.Smith and K.V.Samokhin {\it cond mat/0105109} (unpublished)
\bibitem{Shobo} S. Bhattacharya et al, {\it Phys. Rev.} {\bf B37}, 5901 (1988).
\bibitem{Clean}A. Hosseini et. al., cond-mat/9811041 (unpublished).
\bibitem {Pankert} J. Pankert et. al., {\it Phys. Rev. Lett.}{\bf 65}, 3052 (1990).
\bibitem {Faraday1}D. Dominguez et. al., {\it Phys. Rev. Lett.}{\bf 74}, 2579 (1995).
\bibitem {Faraday2}D. Dominguez et. al., {\it Phys. Rev.}{\bf B53}, 6682 (1996).
\bibitem {Blatter} G. Blatter and B. Ivlev, {\it Phys. Rev.}{\bf B52}, 4588 (1995).
\bibitem {Caroli}C. Caroli, P. G. de Gennes and J. Matricon, {\it Phys. Lett.}{\bf 9}, 307 (1964).
\bibitem {Volovik} G. E. Volovik, {\it JETP Lett.} {\bf 58}, 469 (1993).
\bibitem{STM1}I. Maggio-Aprile et. al., {\it Phys. Rev. Lett.} {\bf 75}, 2574 (1995).
\bibitem{STM2}Ch. Renner et. al., {\it Phys. Rev. Lett.} {\bf 80}, 3606 (1998).
\bibitem{Hirschfeld}C. Kubert and P. Hirschfeld, {\it Solid State Comm.}{\bf 105},
459 (1998).
\bibitem{spht}K. Moller et. al., {\it Phys. Rev. Lett.}{\bf 73}, 2744 (1994).
\bibitem{TVR} T. V. Ramakrishnan and A. K. Rajagopal, {\it Jour. of Stat. Phys.}{\bf 103}, 441 (2001).
\bibitem{DeGennes}P. G. de Gennes  {\it Superconductivity of metals and alloys}
(Addison-Wesley, Reading, MA 1989)
\end{thebibliography}
\end{document}